\documentclass{article}
\usepackage{natbib}
\usepackage[utf8]{inputenc}
\usepackage[T1]{fontenc}    
\usepackage{hyperref}       
\usepackage{url}            
\usepackage{booktabs}       
\usepackage{amsfonts}       
\usepackage{nicefrac}       
\usepackage{microtype}      
\usepackage{lipsum}
\usepackage{comment}

\usepackage{subcaption}
\usepackage{eurosym}
\usepackage{amsmath}
\usepackage[linesnumbered,ruled,vlined]{algorithm2e}
\usepackage{graphicx}
\graphicspath{ {./images/} }

\title{Modelling Solar PV Adoption in Irish Dairy Farms using Agent-Based Modelling}

\author{
 Iias Faiud \\
  School of Computer Science\\
  University of Galway\\
  Galway, Ireland H91 FYH2 \\
  \texttt{i.faiud1@universityofgalway.ie} \\
  \And
  Michael Schukat \\
  School of Computer Science\\
  University of Galway\\
  Galway, Ireland H91 FYH2 \\
  \texttt{michael.schukat@universityofgalway.ie} \\
   \And
 Karl Mason \\
  School of Computer Science\\
  University of Galway\\
  Galway, Ireland H91 FYH2 \\
  \texttt{karl.mason@universityofgalway.ie} \\
}
\begin{document}

\maketitle              

{ 
    \renewcommand{\thefootnote}{\fnsymbol{footnote}}
    \footnotetext{\textit{\textit{*** Proc. of the Deep Learning for Sustainable Precision Agriculture, ECML PKDD} 2023, 22 \textit{September} 2023, \textit{\url{https://sites.google.com/view/dlspa-ecmlpkdd2023/}}. 2023.}}
}

\begin{abstract}

The agricultural sector is facing mounting demands to enhance energy efficiency within farm enterprises, concurrent with a steady escalation in electricity costs. This paper focuses on modelling the adoption rate of photovoltaic (PV) energy within the dairy sector in Ireland. An agent-based modelling approach is introduced to estimate the adoption rate. The model considers grid energy prices, revenue, costs, and maintenance expenses to calculate the probability of PV adoption. The ABM outputs estimate that by year 2022, 2.45\% of dairy farmers have installed PV. This is a 0.45\% difference to the actual PV adoption rate in year 2022. This validates the proposed ABM. The paper demonstrates the increasing interest in PV systems as evidenced by the rate of adoption, shedding light on the potential advantages of PV energy adoption in agriculture. This study possesses the potential to forecast future rates of PV energy adoption among dairy farmers. It establishes a groundwork for further research on predicting and understanding the factors influencing the adoption of renewable energy.

\end{abstract}

\section{Introduction}

Photovoltaic (PV) power generation systems are currently one of the fastest growing in the use of direct solar energy \cite{viveros2020sizing}. To date, PV technology has been employed to supply the required power for various agricultural applications, including water pumping and irrigation, saltwater desalination, crop drying, greenhouse cultivation, etc \cite{gorjian2020farm}. Integrating renewable energy sources into the agricultural and dairy sectors poses a significant challenge. This problem arises due to the increasing demand for energy in these sectors, coupled with the pressing need for sustainable and environmentally friendly practises \cite{breen2020photovoltaic}. Previous systematic reviews of renewable energy sources highlight that the need for more public awareness is a significant barrier to accepting renewable energy technologies \cite{qazi2019towards}. Understanding the factors that drive the adoption of PV energy among farmers is crucial for facilitating the transition to a greener and more sustainable agricultural industry.

There are multiple compelling reasons why integrating renewable energy sources into the agricultural and dairy sectors holds great significance. Firstly, it aligns with the global goal of decarbonizing the energy grid and reducing greenhouse gas emissions \cite{kabeyi2022sustainable}. Several dairy farms already use PV systems to fulfil the electric demands of their equipment and facilities. Implementing solar PV technologies reduces fuel consumption, allowing for the development of more sustainable and flexible technologies \cite{gorjian2020farm}. By transitioning from fossil fuel-based energy to renewable sources such as solar or wind power, these sectors can significantly contribute to mitigating climate change and minimising their environmental footprint \cite{owusu2016review}. Additionally, addressing the integration of renewable energy sources into the agricultural and dairy sectors is crucial for the long-term sustainability of agricultural and dairy operations \cite{erdiwansyah2021critical}. According to a review on incentives for adopting sustainable agricultural practises, the strongest motivation for farmers to adopt sustainable practises is perceived benefits for either their farms, the environment, or both \cite{pineiro2020scoping}.  

Estimating future photovoltaic (PV) adoption rates holds substantial importance for policymakers, as having reliable estimates allows policymakers to develop informed and evidence-based policies and regulations \cite{mayne2018using}. By understanding the anticipated adoption rates, policymakers can design supportive frameworks that incentives farmers to adopt PV systems, such as providing financial incentives or offering streamlined permitting processes. Additionally, accurate estimates of future PV adoption can aid policymakers in long-term energy planning and infrastructure development \cite{maka2022solar}. This information allows them to anticipate the potential increase in electricity generation from PV sources and plan accordingly for grid integration and distribution infrastructure upgrades \cite{nwaigwe2019overview}. Moreover, estimates of PV adoption rates assist policymakers in assessing the potential environmental and economic impacts of widespread PV adoption in the agricultural sector.

Agent-based modelling (ABM) is a computational approach used to simulate the behaviour and interactions of autonomous agents within a complex system \cite{an2021challenges}. This paper focuses on estimating the adoption rate of PV energy among farmers over time using ABM. By simulating the PV adoption process, we aim to gain insights into the dynamics and trends associated with the uptake of this renewable energy source within the dairy sector. The ABM incorporates energy prices and economic considerations.

\par The research presented in this paper makes the following contributions:
\begin{enumerate}
\item To introduce agent-based modelling to dairy farming renewable adoption modelling.
\item To validated the proposed agent-based model by modelling solar PV adoption in the Irish dairy sector.
\end{enumerate}

The outline of the paper is as follows: Section 2 will provide an overview of the relevant literature in agent-based modelling, renewable energy, and simulation. Section 3 will detail the experimental methods employed in the study. The experimental and discussion of the results will be presented in Section 4. Finally, Section 5 will provide conclusions drawn from the study and suggest potential directions for future research.

\section{Related Work}
Agent-based modelling is a powerful simulation technique that has seen several applications in recent years, including applications to real-world problems such as simulating the dynamics of consumer technology adoption \cite{stavrakas2019agent}. In agent-based modelling, a system is modelled as a collection of autonomous decision-making entities called agents. Each agent observes the system’s state and makes decisions based on a set of rules \cite{rai2015agent}.
A previous study that uses agent-based modelling to simulate renewable energy adoption presents an innovative dynamic drought risk adaptation model called ADOPT, this model combines socio-hydrological and agent-based modelling approaches to evaluate the factors that influence adaptation decisions and the subsequent adoption of measures and how this affects drought risk for agricultural production. The authors compare adaptive behavioural theories, including the protection motivation theory, which describes bounded rationality, with business-as-usual and economic rational adaptive behaviour. The results show that an agent-based approach can improve estimations of drought risk and the need for emergency food aid. \cite{wens2020simulating}.

An additional study provides an overview of agent-based modelling and its applications, including flow simulation, organisational simulation, market simulation, and diffusion simulation. The author argues that agent-based modelling is a mindset more than a technology: it describes a system from the perspective of its constituent units \cite{bonabeau2002agent}.

In the study by Mason et al., an agent-based methodology is employed to analyse the prospects of long-term investment within the power generation sector in Great Britain, with particular emphasis on ensuring the operational viability of the system. The research primarily aims to evaluate the effectiveness of various policies to diminish emissions and foster the deployment of renewable generation and battery storage technologies. The findings derived from the modelling exercise indicate that, although the expense associated with battery storage is anticipated to decline progressively over time, a significant subsidy is still imperative to validate the economic rationale behind investing in this technology \cite{mason2021investing}.

A previous study by Meles et al. is modelling the adoption of heating systems in Ireland. The authors present an agent-based model to simulate the adoption of heat pumps in Ireland. The model incorporates various factors influencing adoption decisions, including expected energy savings, government incentives, and social influence. The authors use the model to explore different scenarios and evaluate the effect of other policy interventions on the adoption rate of heat pumps. The results show that government incentives, such as grants and subsidies, can significantly increase the adoption rate of heat pumps. In addition, social influence plays a vital role in shaping adoption decisions: households are more likely to adopt heat pumps if their neighbours have already done so. Overall, this paper provides valuable insights into the factors that drive the adoption of renewable home heating systems and the potential impact of policy interventions. Using an agent-based modelling approach, the authors can capture the complex dynamics of individual decision-making and explore the potential outcomes of different scenarios \cite{meles2022adoption}.

To date, there have been no applications of ABM to model solar PV adoption in dairy farming in the literature. The simulation-based predictions can inform policymakers and stakeholders in developing targeted strategies to accelerate the adoption of PV energy in the agricultural sector and drive sustainable agricultural practises.

\section{Methodology}

Our work employs an agent-based modelling methodology to predict the adoption rate of photovoltaic (PV) energy among dairy farmers. Agent-based modelling enables the representation of individual decision-making entities, referred to as agents, within a simulation framework.
The proposed method aims to simulate the adoption of (PV) systems by farmers, represented by agents in our model, over a designated time period. The methodology involves the development of an iterative algorithm that captures the decision-making process of dairy farm owners considering various economic factors. Input parameters such as PV cost, maintenance cost, discount rate, energy price, total farmers, subsidies, and the starting year of the simulation are provided. As shown below, the algorithm calculates the economic utility based on these parameters, representing the attractiveness of PV adoption. Probability calculation is performed using function that consider economic utility and determine the likelihood and proportion of PV adoption. The energy price is updated during each iteration by considering historical data \cite{SEAI} in order to accommodate possible changes over time. The algorithm progresses by incrementing the year until the predefined end year is reached. The simulation results provide the number of PV systems adopted in dairy farms in 2022.

\begin{table}[h]
\caption{Model inputs}
\centering
\begin{tabular}{l@{\hskip 1in}l}
\toprule
Input data          & Value         \\ 
\midrule
PV cost        & 5000-15000 \EUR{}     \\ 
Maintenance cost    & 2\% of PV cost \\ 
Discount rate       & 4\%            \\ 
Total farmers       &18000  \\
Subsidies      & 1000-3500 \EUR{}     \\ 
Year                & 2005-2022      \\

\bottomrule
\end{tabular}
\end{table}

\begin{algorithm}
\caption{Agent-Based Model for Dairy Farm PV Adoption}
\SetAlgoLined
\SetKwInOut{Input}{Input}
\SetKwInOut{Output}{Output}

\Input{Initial values: Pv\_cost, Maintenance\_cost, Discount\_rate, Energy\_price, Total\_farmers, Subsidies, Year}
\Output{Pv adoption over time}
\While{simulation not ended}{
    $Economic\_utility \leftarrow \text{EU} = \text{NPV} - \text{IIC} + \text{Sub}$\;
    $Probability \leftarrow \frac{\beta}{1 + \exp\left(-\alpha \cdot \frac{\text{economic utility}}{\text{total farmers}}\right)}$\;
    $PV\_adoption \leftarrow Probability \times Total\_farmers$\;
    $Energy\_price \leftarrow \text{Get the next energy value based on the year}(Year)$\;
    $Subsidies \leftarrow \text{Get the next subsidy value based on the year}(Year)$\;
    $Year \leftarrow Year + 1$\;
}
\end{algorithm}

The economic utility, a key measure in evaluating the viability of solar systems, can be computed using equation (1), which accounts for various factors.
\[
{\text{Economic utility}} = {\text{NPV}} - {\text{IIC}} + {\text{Subsidies}} \tag{1}
\]
The first component, the Net present value of Energy savings, represented in equation (2), It takes into account the energy savings over time ${R_t}$ and the discount rate ${i}$. The equation sums up the discounted energy savings for each period (t) to determine the overall net present value. The second component, the Initial investment cost, encompasses the expenses incurred during the installation of the solar system. Lastly, Subsidies denote the financial incentives or support provided by external entities to alleviate the installation cost of the solar system.
\[
\text{Net present value} = \sum_{t=0}^{n} \frac{R_t}{{(1 + i)}^t} \tag{2}
\]

The probability of adaptation represents the likelihood of photovoltaic (PV) system adoption by farmers. It calculates the probability based on several factors, including $\beta$ value, economic utility, $\alpha$, and the total number of farmers, as shown in equation (3).
\[
\text{Probability} = \frac{\beta}{1 + \exp\left(-\alpha \cdot \frac{\text{economic utility}}{\text{total farmers}}\right)} \tag{3}
\]

The $\beta$ represents the initial likelihood of adoption in the absence of specific influencing factors, while the $\alpha$ determines the rate of change in the adoption probability in response to changes in economic utility. This equation makes it possible to assess the likelihood of PV system adoption among dairy farmers.


\begin{figure}
  \centering
  \includegraphics[width=0.95\textwidth]{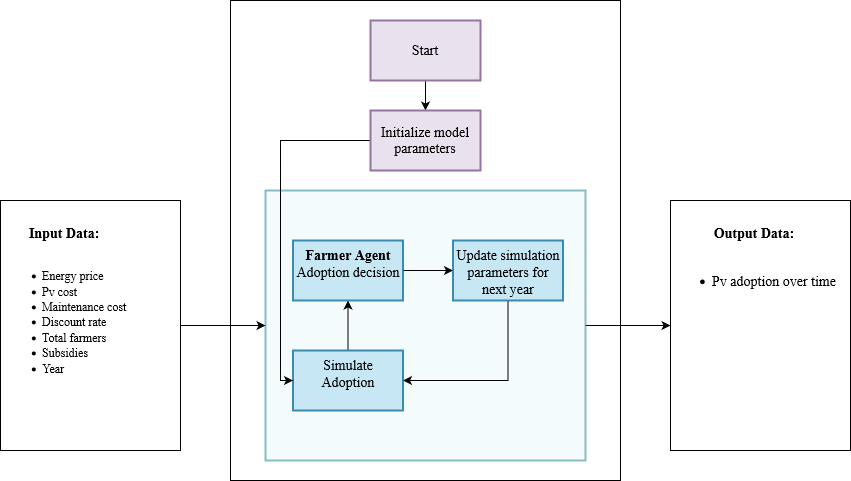}
  \caption{The structure of the agent-based model for the PV system adoption.}
  \label{fig:image}
\end{figure}

\section{Experimental Results}

The results obtained from the ABM model's projection of PV adoption closely align with the number of solar PV systems in 2022. The model's output indicated that out of the 18,000 farmers considered in the simulation, 441 farmers adopted PV systems in 2022. This adoption rate represents a minor overestimation of 0.45\% compared to the actual adoption figures. The actual number of farmers reported to have adopted PV in Ireland is 360 in 2022 \cite{irishexaminer}.

This close correspondence between the model's projected adoption rate and the actual number of PV systems in 2022 validates the accuracy and effectiveness of the model. It demonstrates that the model's predictions are reliable in capturing real-world adoption patterns within the dairy sector. The ability of the model to replicate the observed adoption rate showcases its capability to provide valuable insights and predictions regarding PV adoption trends in agricultural settings.

These findings confirm the accuracy of the adopted modelling approach and underscore its efficacy in studying and forecasting the future of PV adoption in the dairy sector. By successfully capturing the adoption patterns and closely aligning them with the actual number of PV systems in 2022, the model demonstrates its potential for providing valuable insights into the future trajectory of PV adoption within the dairy industry. With its ability to simulate and project adoption rates over time, the modelling approach becomes an indispensable tool for policymakers, researchers, and industry stakeholders. By considering various factors such as PV cost, subsidies, energy prices, and other influential variables, the model can generate reliable forecasts that inform decision-making and strategic planning related to PV adoption in the dairy sector. By leveraging this modelling approach, stakeholders can gain a deeper understanding of the dynamics that drive PV adoption trends within the dairy industry. They can anticipate future adoption rates, identify potential barriers or opportunities, and formulate targeted interventions to accelerate or optimise PV integration in agricultural settings.

\begin{figure}
  \centering
  \includegraphics[width=0.95\textwidth]{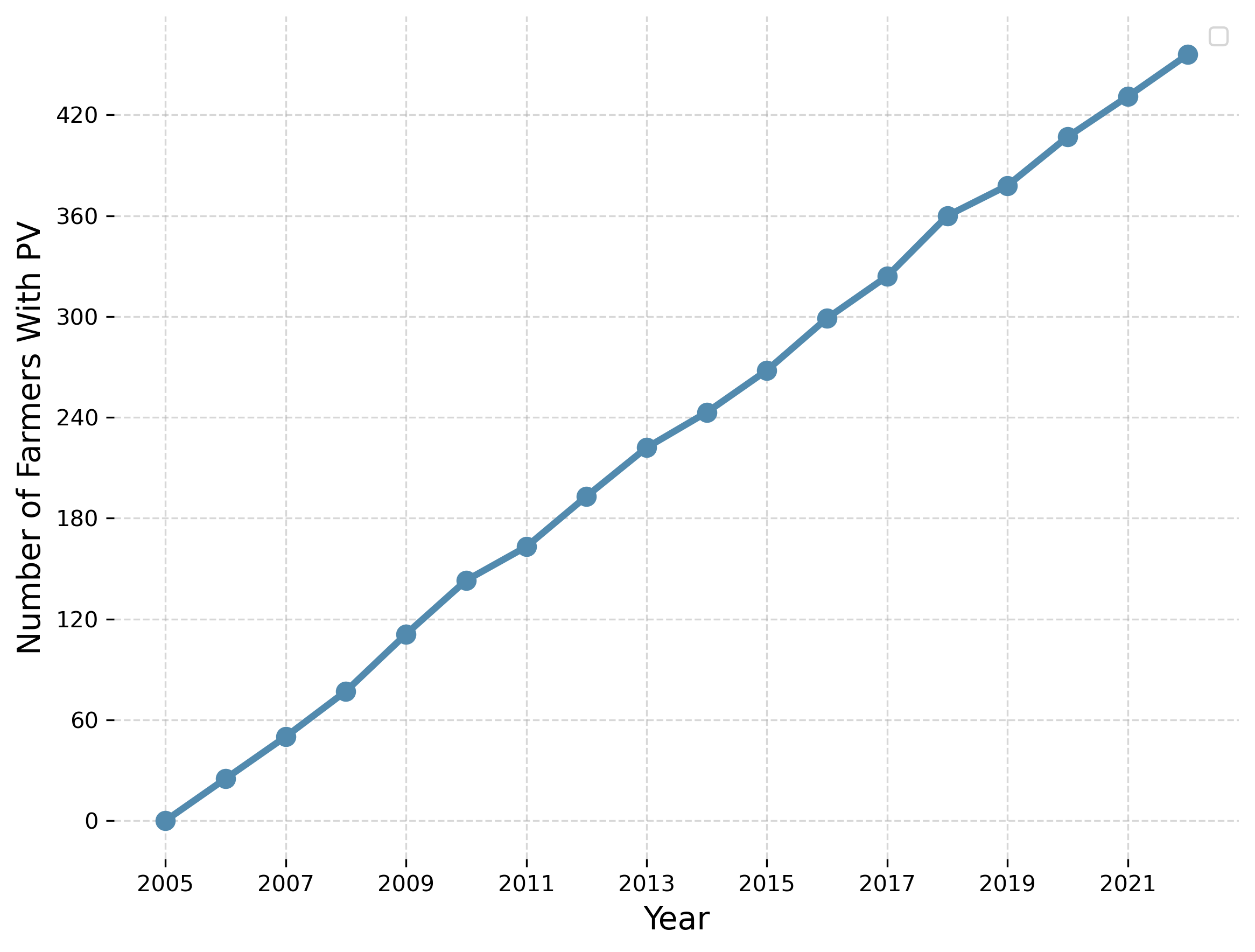}
  \caption{ABM Output of Dairy Farm Solar PV Adoption Over Time.}
  \label{fig:image}
\end{figure}

\section{Conclusion and Future Work}

This study employed an agent-based modelling approach to predict the adoption rate of photovoltaic (PV) energy among dairy farmers. The model was validated between 2005 and 2022. The ABMs estimate of the number of dairy farms with PV in 2022 is 0.45\% different to the actual number of dairy farms recorded to have PV in 2022. The methodology involved an iterative algorithm that captured the decision-making process of dairy farm owners considering economic factors.

\begin{enumerate}
\item Introduced an agent-based model to model the adoption of renewable energy in the dairy farming sector in Ireland.
\item Validated the agent-based model by comparing its results with real-world data on solar PV adoption in the Irish dairy sector.

\item The adopted modelling approach is accurate and effective, and it can be used to study and forecast the future of PV adoption in the Irish dairy sector.

\end{enumerate}

Future work will focus on refining the simulation model by modelling other renewable resources and modelling policy makers as agents, enabling a better understanding of PV adoption in the dairy farming sector.

\section*{Acknowledgements} This publication has emanated from research conducted with the financial support of Science Foundation Ireland under Grant number [21/FFP-A/9040].

%
%
%
 \bibliography{template}

\end{document}